\documentclass[conference]{IEEEtran}
\IEEEoverridecommandlockouts
\usepackage{cite}
\usepackage{amsmath,amssymb,amsfonts}
\usepackage{algorithmic}
\usepackage{graphicx}
\usepackage{textcomp}
\usepackage{xcolor}
\def\BibTeX{{\rm B\kern-.05em{\sc i\kern-.025em b}\kern-.08em
    T\kern-.1667em\lower.7ex\hbox{E}\kern-.125emX}}

\makeatletter
\def\ps@IEEEtitlepagestyle{%
	\def\@oddfoot{\mycopyrightnotice}%
	\def\@oddhead{\hbox{}\@IEEEheaderstyle\leftmark\hfil\thepage}\relax
	\def\@evenhead{\@IEEEheaderstyle\thepage\hfil\leftmark\hbox{}}\relax
	\def\@evenfoot{}%
}
\def\mycopyrightnotice{%
	\begin{minipage}{\textwidth}
		\centering \scriptsize
		Copyright~\copyright~20XX IEEE.  Personal use of this material is permitted.  Permission from IEEE must be obtained for all other uses, in any current or future media, including reprinting/republishing this material for advertising or promotional purposes, creating new collective works, for resale or redistribution to servers or lists, or reuse of any copyrighted component of this work in other works.
	\end{minipage}
}
\makeatother

\usepackage{changes}
\usepackage{mathtools}
\usepackage{subfig}
\usepackage{algorithm2e}
\usepackage{amsmath}
\SetKwInput{KwInput}{Input}
\SetKwInput{KwOutput}{Output}
\usepackage{comment}
\DeclareMathOperator{\Tr}{Tr}

\begin{document}
\DeclarePairedDelimiter\ceil{\lceil}{\rceil}

\title{Calibration-Aware Transpilation for Variational Quantum Optimization\\
}

\author{\IEEEauthorblockN{Yanjun Ji}
	\IEEEauthorblockA{\textit{Institute of Computer Architecture}\\ \textit{and Computer Engineering} \\
		\textit{University of Stuttgart}\\
		Stuttgart, Germany \\
		yanjun.ji@informatik.uni-stuttgart.de}
	\and
	\IEEEauthorblockN{Sebastian Brandhofer}
	\IEEEauthorblockA{\textit{Institute of Computer Architecture}\\ \textit{and Computer Engineering} \\
		\textit{University of Stuttgart}\\
		Stuttgart, Germany \\
		sebastian.brandhofer@informatik.uni-stuttgart.de}
	\and
	\IEEEauthorblockN{Ilia Polian}
	\IEEEauthorblockA{\textit{Institute of Computer Architecture}\\ \textit{and Computer Engineering} \\
		\textit{University of Stuttgart}\\
		Stuttgart, Germany \\
		ilia.polian@informatik.uni-stuttgart.de}
}

\maketitle

\begin{abstract}
Today's Noisy Intermediate-Scale Quantum (NISQ) computers support only limited sets of available quantum gates and restricted connectivity. Therefore, quantum algorithms must be transpiled in order to become executable on a given NISQ computer; transpilation is a complex and computationally heavy process. Moreover, NISQ computers are affected by noise that changes over time, and periodic calibration provides relevant error rates that should be considered during transpilation.
Variational algorithms, which form one main class of computations on NISQ platforms, produce a number of similar yet not identical quantum ``ansatz'' circuits. In this work, we present a transpilation methodology optimized for variational algorithms under potentially changing error rates. We divide transpilation into three steps: (1) noise-unaware and computationally heavy pre-transpilation; (2) fast noise-aware matching; and (3) fast decomposition followed by heuristic optimization. For a complete run of a variational algorithm under constant error rates, only step (3) needs to be executed for each new ansatz circuit. Step (2) is required only if the error rates reported by calibration have changed significantly since the beginning of the computation. The most expensive Step (1) is executed only once for the whole run. This distribution is helpful for incremental, calibration-aware transpilation when the variational algorithm adapts its own execution to changing error rates. Experimental results on IBM's quantum computer show the low latency and robust results obtained by calibration-aware transpilation.
\end{abstract}


\section{Introduction}

Quantum computing promises fundamentally more efficient solutions for a number of hard real-world problems. In the current Noisy Intermediate-Scale Quantum (NISQ) era, variational algorithms \cite{vqa} such as the Quantum Approximation Optimization Algorithm (QAOA) \cite{farhi2014quantum, farhi2016quantum,Choi19, zhu2020adaptive, zhou2020quantum,HadfieldWORVB19, Fernandez-Pendas22} or Variational Quantum Eigensolver (VQE) \cite{peruzzo2014variational, mcclean2016theory, romero2018strategies} are receiving significant attention, since they can cope with non-trivial error rates of NISQ computers. Variational algorithms interchange classical and quantum computations. One complete run of a variational algorithm executes a number of quantum ``ansatz'' circuits that are parameterized, i.e., have identical basic structure but differ in some specific parameters.

\begin{figure}[b!]
	\centerline{\includegraphics[width=\columnwidth]{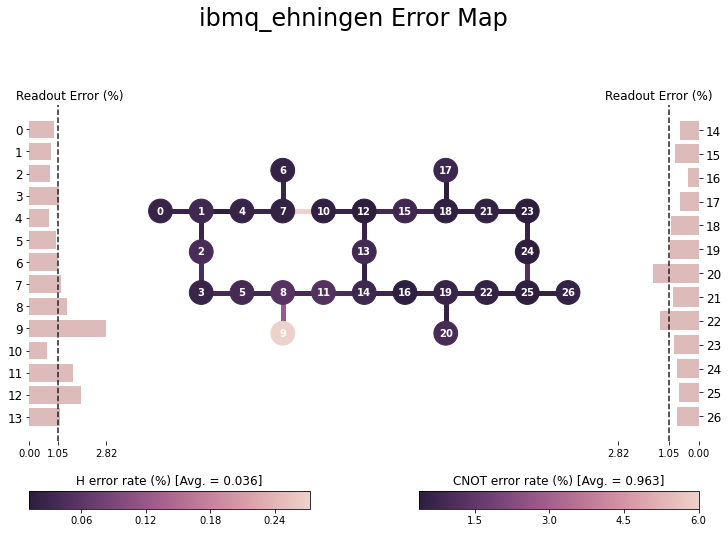}}
	\caption{Topology graph of {\em ibmq\_ehningen} with error information of single-qubit and two-qubit gates and readout.}
	\label{fig:error_map}
\end{figure}

State-of-the-art NISQ computers come with limitations with respect to connectivity of their qubits and quantum operations supported. Moreover, they are affected by comparatively high noise levels that can strongly vary over time. For example, computers that are part of IBM Quantum Experience (IBM QX) undergo an hourly calibration, which includes error characterization, and the determined error rates are provided to their users. To illustrate the role of calibration, Fig.~\ref{fig:error_map} shows the topology graph of the 27-qubit IBM QX machine {\em ibmq\_ehningen} along with a snapshot of error rates for its components. It includes error rates for each single-qubit gate, all two-qubit (\texttt{CNOT} or \texttt{cx}) gates between qubits connected according to topology graph, and for readout operations.

\begin{figure*}[tb]
	\centerline{\includegraphics[width=\linewidth]{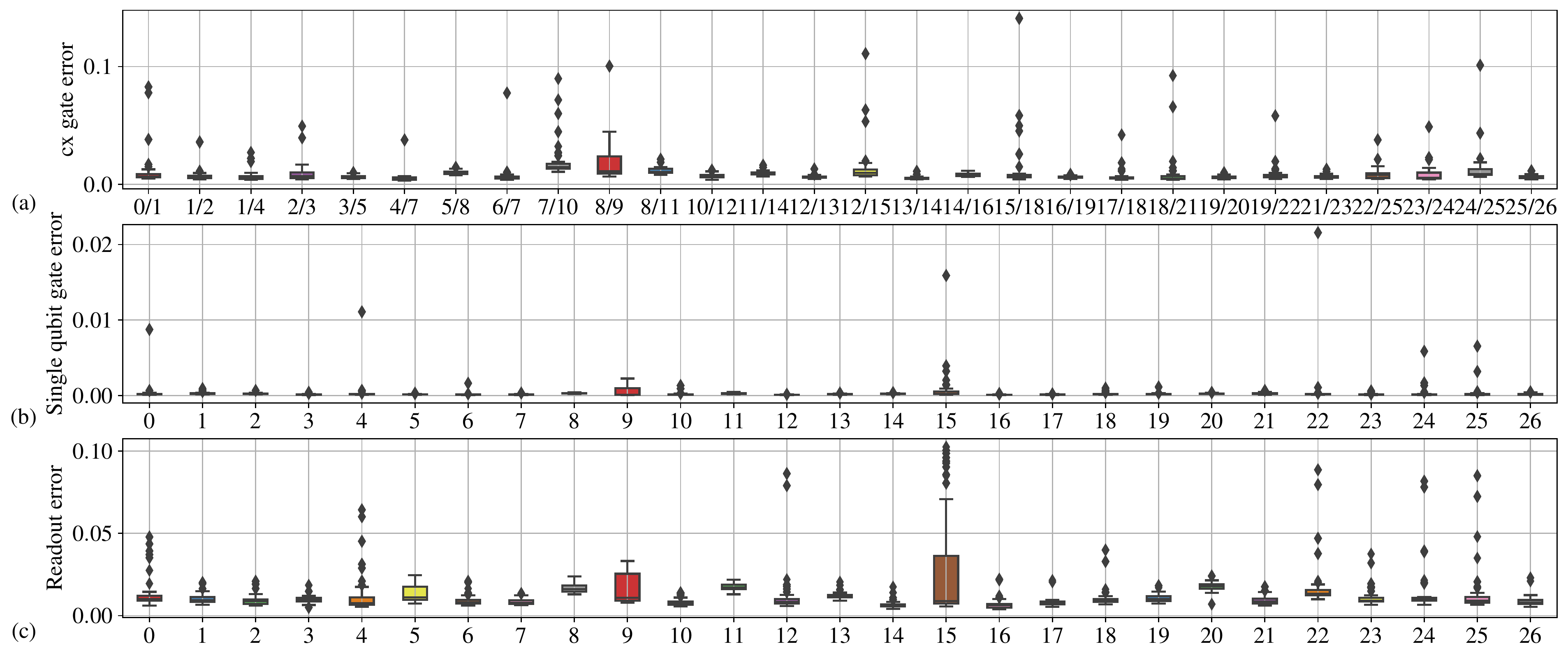}}
	\caption{Error information of {\em ibmq\_ehningen} over 39 days.}
	\label{fig:error_data}
\end{figure*}

We can see that error rates differ widely both across classes of errors (e.g., they are an order of magnitude higher for readout and two-qubit operations than for single-qubit gates) and also within one class. Hourly calibration data, which we collected on {\em ibmq\_ehningen} over a period of 39 days, are reported in Fig.~\ref{fig:error_data} and expose large-scale fluctuation in the temporal domain as well. The highest variations were observed for the \texttt{cx} gate between qubits $8$ and $9$ (Fig.~\ref{fig:error_data}a) and for the readout errors of qubit $15$ (Fig.~\ref{fig:error_data}c).

In general, quantum circuits, including the ansatz circuits of variational algorithms, use operations and qubit interactions that are not supported by a given NISQ architecture. For this reason, they need to be transpiled: all their operations must be mapped to that architecture's quantum gates, and two-qubit gates must either be mapped to connected qubits or proximity must be established by adding \texttt{swap} gates. In addition, transpilation should be noise-aware, that is, try to use qubits with currently lowest error rates. Various transpilation \cite{Sivarajah2020tketAR,amy2020staq,harrigan2021quantum,steiger2018projectq,murali2019full,KhammassiASNKRL22,PennyLane} and heuristic \cite{SSCP:19, SiraichiSCP18, ZulehnerPW19, Murali19, TannuQ19, LiDX19, childs19, LiMZY20, Niu20} and exact \cite{SiraichiSCP18,BhattacharjeeC17,ShafaeiSP14,Pedram16,Murali19,TanC20} mapping methods have been proposed. Both transpilation and mapping are considered to be computationally hard problems.

\begin{figure}[t]
	\centerline{\includegraphics[width=\linewidth]{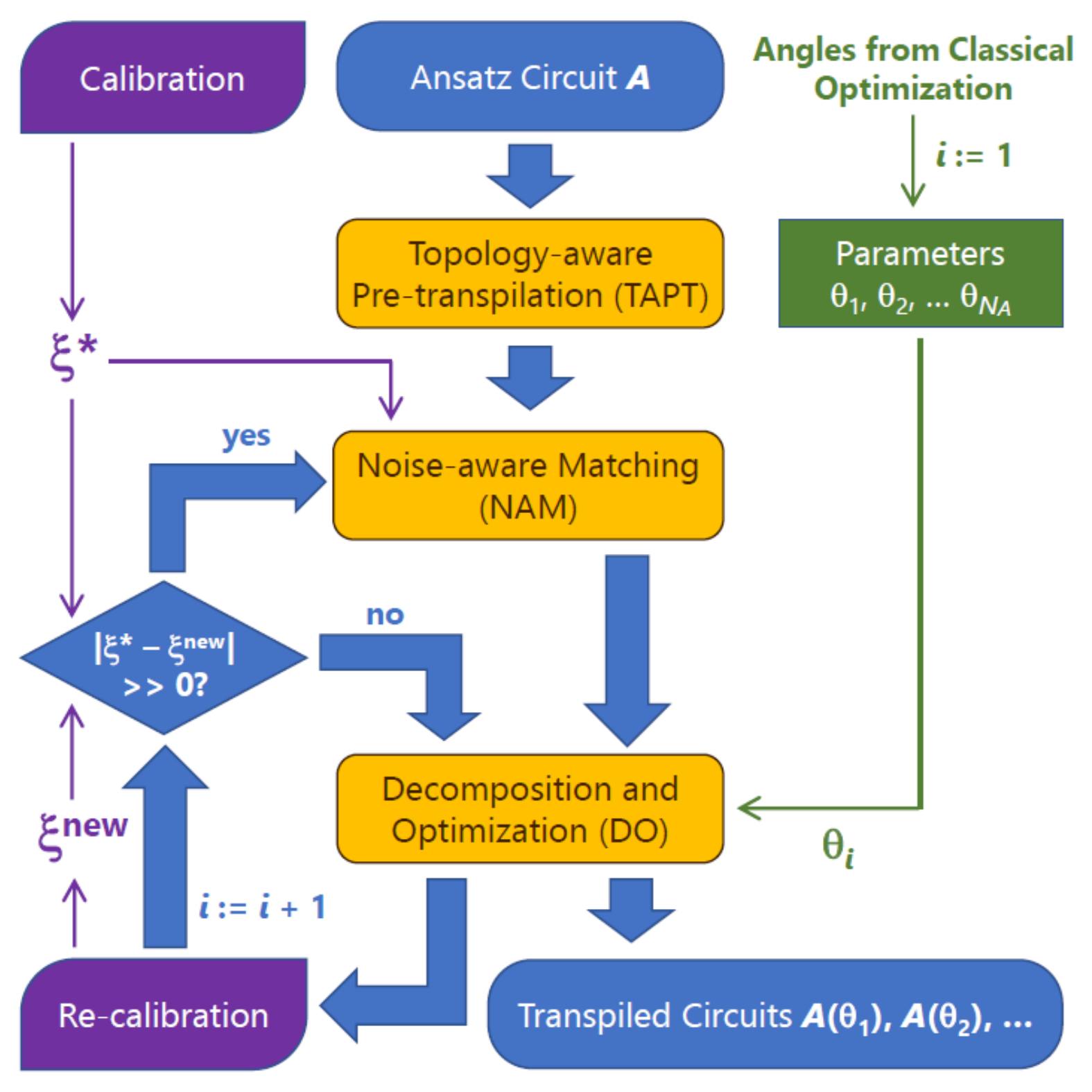}}
	\caption{Flowchart of Calibration-Aware (CA) Transpilation.}
	\label{fig:flow}
\end{figure}

In this paper, we introduce calibration-aware transpilation, which is optimized for variational algorithms with parameterized ansatz circuits and avoids the need for a costly complete transpilation of each new circuit. The procedure is outlined in Fig.~\ref{fig:flow}. The processed ansatz circuit $A$ is parameterized with values $\theta_i$, which, in case of QAOA, are angles determined by the classical optimization step. Transpiled circuits $A(\theta_1), A(\theta_2), \ldots$ differ only minimally, and their basic structure must be computed only once for $A$ and can be reused by all circuits. Moreover, transpilation takes error rates $\xi^*$ into account, which are determined by calibration.

After each re-calibration, the procedure checks whether new error rates $\xi^{\rm new}$ are substantially different from $\xi^*$. If this is the case, transpilation is not repeated from scratch, but the initial, optimized solution is mapped to a different subset of qubits with the same sub-graph topology yet better fidelities. Overall, calibration-aware transpilation consists of three steps:
Topology-Aware Pre-Transpilation (TAPT), executed once for the entire run of a variational algorithm and calculating parts of the solution applicable to all ansatz circuits; Noise-Aware Matching (NAM), invoked only when error rates have changed significantly; and Decomposition and Optimization (DO), which includes improvements for a specific ansatz circuit $A(\theta_i)$.

The main advantage of calibration-aware transpilation is the significantly reduced effort, as all computationally heavy parts are accumulated in the TAPT step that is run only once, and the two remaining steps are lightweight.
It is feasible to use NAM and DO in an incremental mode: whenever a new ansatz circuit is ready for execution on the quantum hardware, reserve the quantum computer, acquire calibration data, execute NAM (if needed) and DO, and then immediately execute the transpiled circuit on the reserved computer. This makes sure that the most recent calibration data are used for each ansatz circuit, while the amount of the quantum computer's time ``wasted'' during reservation is minimal. In the traditional approach with full transpilation of each ansatz circuit, reserving the quantum computer for its complete duration would be unrealistic. The computer would start processing other tasks, and the transpiled ansatz circuit, once it is ready, would be inserted into the regular queue and executed possibly at a time instant when the calibration data is outdated.

\begin{figure}[t]
	\centerline{\includegraphics[width=\columnwidth]{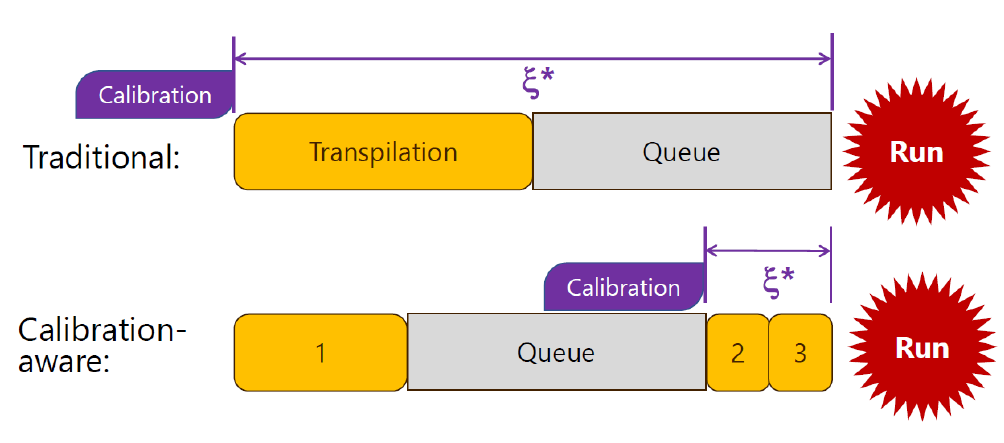}}
	\caption{Advantage for transpiling one circuit with calibration-aware transpilation compared to traditional: fresher error rate information. 1, 2 and 3 indicate TAPT, NAM and DO processes in CA transpilation, respectively.}
	\label{fig:ca-onecircuit}
\end{figure}

\begin{figure*}[htbp]
	\centerline{\includegraphics[width=\linewidth]{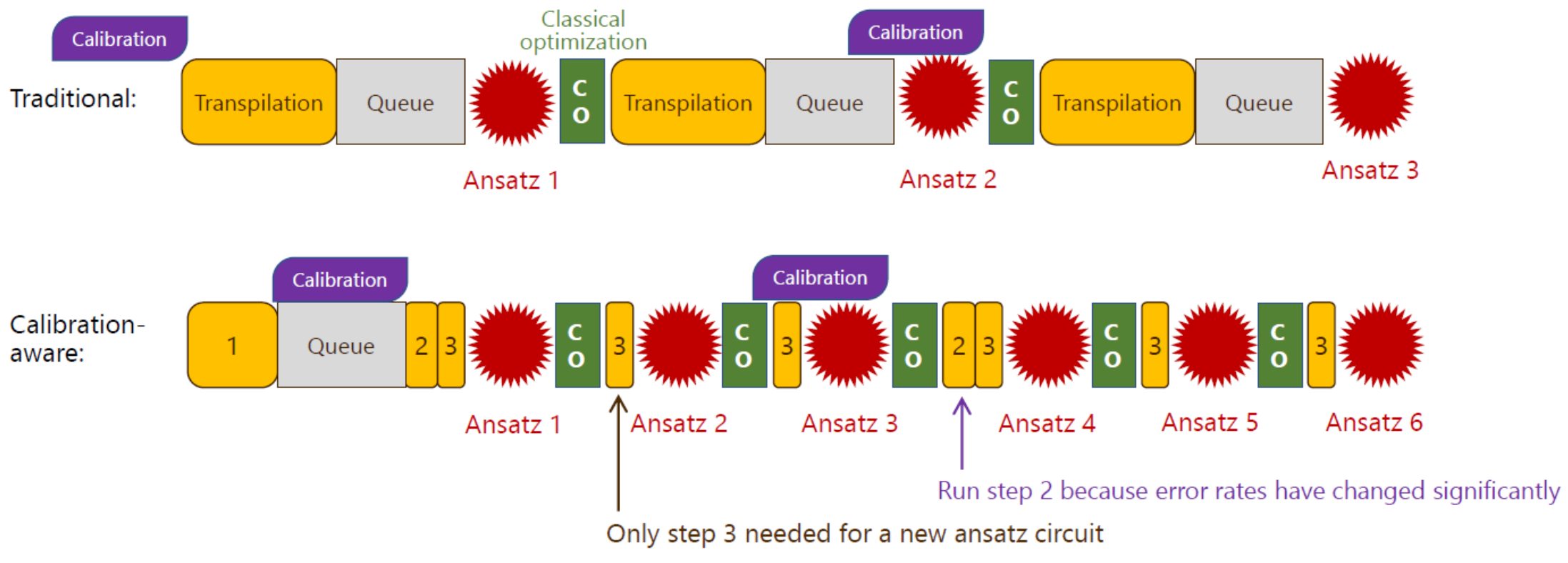}}
	\caption{Advantage of calibration-aware transpilation compared to traditional for a variational run with multiple ansatz circuits: effort reducing and time saving. Classical optimization is used to calculate new parameters of QAOA.}
	\label{fig:ca-multicircuit}
\end{figure*}

In addition, a potential advantage of transpiling one circuit with calibration-aware transpilation is shown in Fig.~\ref{fig:ca-onecircuit}. The high-level idea is to submit the circuit directly to the queue after step 1, which is the most computationally heavy process. Steps 2 and 3, which take calibration data into account, are performed just before 
the circuit is due for 
execution, followed by immediate execution of the transpiled circuit. This ensures that CA has a fresher error rate information than the traditional approach where calibration data is acquired at the beginning of transpilation. This is crucial to the performance of algorithms on the NISQ computers as their errors change over time. Furthermore, based on CA's structures, we can significantly reduce effort and save time for a variational run with multiple circuits, see Fig.~\ref{fig:ca-multicircuit}. While the traditional approach requires each circuit to be transpiled and passed into the queue before execution, CA only needs to pass the first circuit into the queue, and since steps 2 and 3 are fast, the remaining ansatz circuits can be run in one piece, resulting in a significant reduction in overheads.

The remainder of the paper is organized as follows. The next section reviews variational quantum algorithms with a focus on QAOA and includes some investigations of its behavior under noise using simulations. Section \ref{sec:ca} provides details on the individual steps of calibration-aware transpilation. Section \ref{sec:expres} reports results of calibration-aware transpilation in comparison with other methods on several physical quantum computers, outlining its advantages in both: solution quality and runtime. Section \ref{sec:concl} concludes the paper.

\section{Variational Quantum Optimization}
\label{sec:var}

\subsection{Quantum Approximation Optimization Algorithm (QAOA)}

Using the quantum approximate optimization algorithm (QAOA), approximate solutions to computationally hard problems such as portfolio optimization can be computed. QAOA repeatedly performs two alternating steps.
First, a set of parameters is chosen that are used to construct a parameterizable quantum circuit called ansatz circuit. The second step consists of the execution of such an ansatz circuit on a quantum computer to yield a set of measurement results that are evaluated subject to a problem-specific objective function. Then, again a set of parameters is chosen by a classical optimizer that uses the values of the previous objective functions and/or the gradient of that objective function to determine the next set of parameters. These two steps are repeated until the value of the objective function converges or the runtime budget is depleted. 

\begin{figure}[t]
	\centerline{\includegraphics[width=\columnwidth]{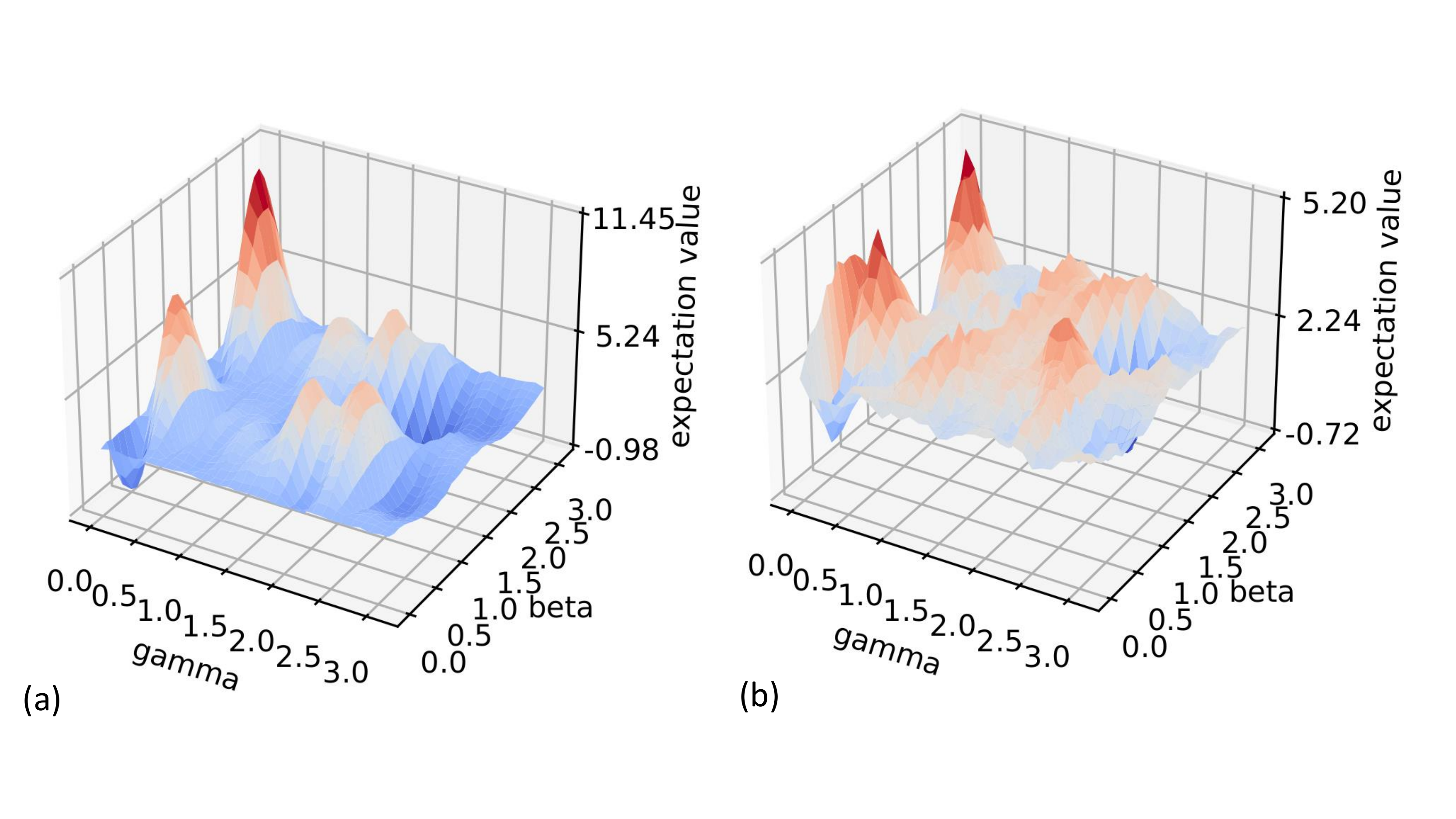}}
	\caption{Expectation values of QAOA with (a) Qasm simulator and (b) {\em ibmq\_ehningen}.}
	\label{fig:surface}
\end{figure}

We use QAOA for portfolio optimization to evaluate the quality of transpilation. Assume $n$ and $B$ are the number of available assets and the number of assets to be chosen, respectively. For each $i \in \{1, ..., n\}$, we introduce variables $z_{i} \in \{0, 1\}$ indicating whether this stock is picked or not. Approximation ratio (AR) is defined as:
\begin{equation}
{\rm AR}(z_1,\dots,z_n)=\left\{\begin{array}{cc} \frac{F(z_1,\dots,z_n)-F_{\rm max}}{F_{\rm min}-F_{\rm max}} & \text{if }\sum_i z_i = B\\
0 & \text{if }\sum_i z_i \neq B\end{array}\right.
\end{equation}
with $F$ the cost function \cite{hodson2019portfolio}.
Success probability is defined as the the probability of obtaining the optimal portfolio. We used $n=5$, $B=2$ with QAOA depth $p = 1$ in our experiments, i.e. the QAOA circuit has 5 qubits. Its depth is 19 and the total number of gates is 50, including 20 \texttt{cx} gates and 5 measurement gates.

The performance of QAOA depends on the initial values. Optimal parameters result in a better performance, i.e., a higher value of approximation ratio and/or success probability. The initial values of QAOA are usually obtained by classical optimizer that finds a local minimum in the area of attraction around the initial point of the probe. The optimal initial values of QAOA with $p=1$ can be determined by grid search. Fig.~\ref{fig:surface} (a) and (b) show the optimization landscape of QAOA using qasm simulator in absence of noise and physical quantum computer {\em ibmq\_ehningen}, respectively. The optimal initial values $\theta$, expectation values $E$, as well as approximation ratio and success probability are shown in Table \ref{tab:optimal_inital_values}. We can see that the optimal initial values of QAOA and optimization landscape are hardly changed, i.e. this QAOA circuit is noise-tolerant and its parameters optimization is barely affected by noise.

\begin{table}[tb]
	\caption{Optimal solution founded by grid search with qasm simulator in absence of errors (error-free) and {\em ibmq\_ehningen}. $\theta$: initial values. $E$: expectation values.}
	\begin{center}
		\begin{tabular}{|c|c|c|c|}
			\hline
			\textbf{} & \textbf{\textit{error-free}}& \textbf{\textit{ ibmq\_ehningen}}\\
			\hline
			$\theta$& $(2.3, 2.1)$&$(2.3, 2.3)$ \\
			\hline
			$E$& $-0.957$&$-0.726$ \\
			\hline
			AR& $0.417$&$0.409$ \\
			\hline
			SP& $0.235$&$0.295$ \\
			\hline
		\end{tabular}
		\label{tab:optimal_inital_values}
	\end{center}
\end{table}

\subsection{Simulation of QAOA with Noise Model}\label{Sim}

\begin{figure}[b]
	\centerline{\includegraphics[width=\linewidth]{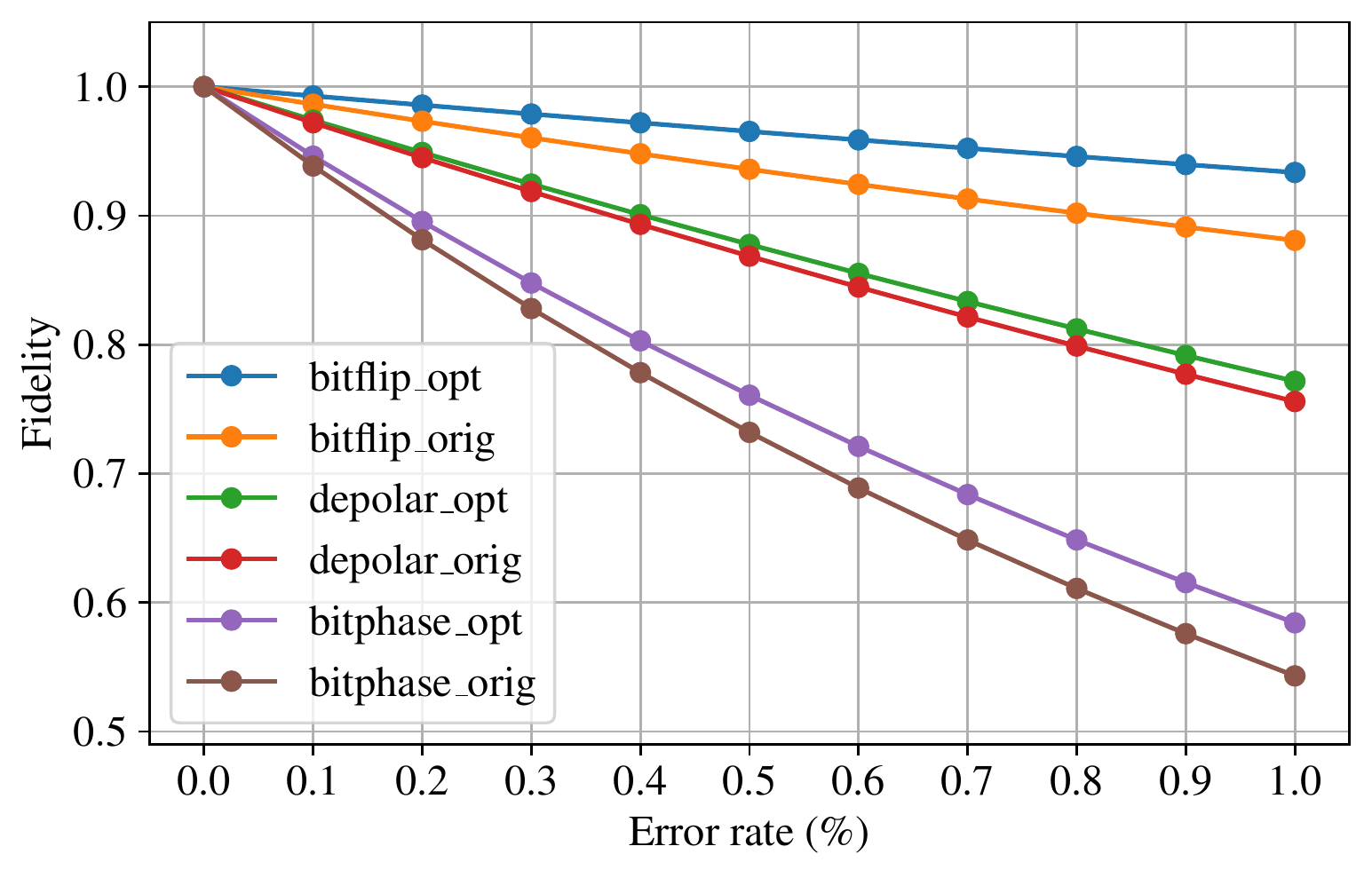}}
	\caption{Simulation of fidelity of QAOA with original (orig) and optimal (opt) initial values using qasm simulator under bitflip, depolarizing and bit-phase flip errors as a function of error rate from $0$ to $1\%$.}
	\label{fig:noise_simulation_fid}
\end{figure}

\begin{figure*}[tb]
	\centerline{\includegraphics[width=\linewidth]{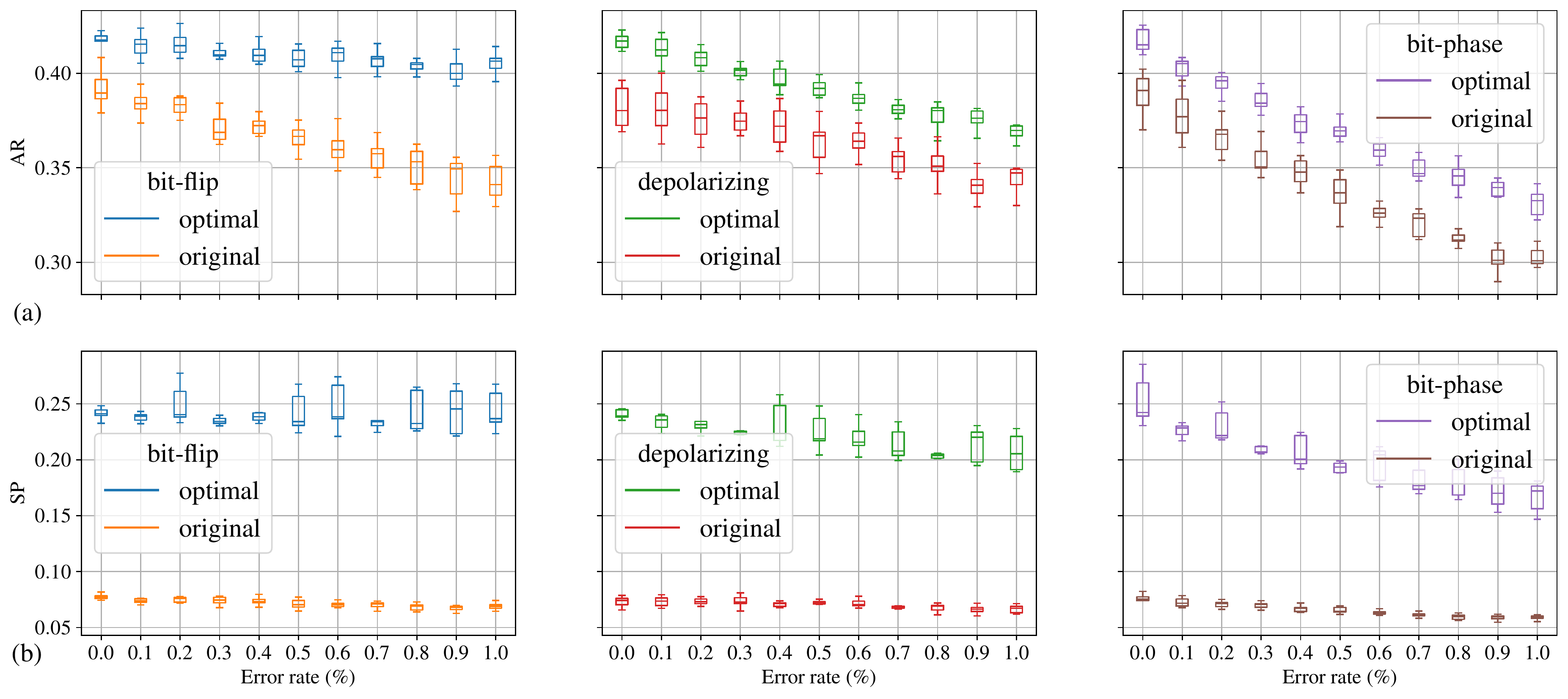}}
	\caption{Simulation of approximation ratio (a) and success probability (b) of QAOA with 10 repetitions using qasm simulator under bitflip, depolarizing and bit-phase flip errors as a function of error rate from $0$ to $1\%$.}
	\label{fig:noise_simulation}
\end{figure*}

We simulate QAOA with original and optimal initial values using bit-flip, bit-phase flip and depolarizing error channels indicated by $\mathcal{E}_X$, $\mathcal{E}_Y$ and $\mathcal{E}_D$ \cite{Nielsen16} with error rate $\lambda$. The error channels act on qubits described by density matrix $\rho$ are defined as:
\begin{align*}
\mathcal{E}_X(\rho) &= \lambda X \rho X+ (1-\lambda)\rho\\
\mathcal{E}_Y(\rho) &= \lambda Y \rho Y+ (1-\lambda)\rho\\
\mathcal{E}_D(\rho) &= \frac{\lambda}{4}\left(X\rho X + Y\rho Y + Z\rho Z\right)+(1 - \frac{3\lambda}{4})\rho.
\end{align*}
We study state fidelity of QAOA final state with error rate up to $1\%$. Moreover, the behavior of approximation ratio and success probability with increased error rates is investigated.

The state fidelity of two quantum states is defined as
\begin{equation}
    F(\rho_1 \rho_2) = \Tr\left[\sqrt{\sqrt{\rho_1}\rho_2\sqrt{\rho_1}}\right]^2
\end{equation}
where $\rho_1$ and $\rho_2$ are density matrices of two quantum states. In our case, $\rho_1$ is the final QAOA state in absence of errors and $\rho_2$ is the state for QAOA with error rate $\lambda$. We consider discrete values $\lambda \in \{0, 0.1\%, 0.2\%, ..., 1\%\}$. The maximum fidelity $1$ occurs at $\lambda = 0$.

In Fig.~\ref{fig:surface}(a), ``orig'' stands for original initial values computed by the classical optimizer COBYLA and ``opt'' denotes optimal initial values determined by grid search.
As shown in Fig.~\ref{fig:noise_simulation_fid}, optimal initial values of QAOA produce a better fidelity than original. The effect of bit-phase flips on fidelity is significant, while the fidelity under bit-flip errors varies only slightly. In addition, the type of initial values produces only a small difference under depolarizing error. At $\lambda = 1\%$, the fidelity of QAOA with bit-flip, depolarizing and bit-phase flip errors drops to about $0.95$, $0.75$ and $0.6$, respectively.

The approximation ratio and success probability of 10 QAOA runs with qasm simulator are shown in Fig.~\ref{fig:noise_simulation} (a) and (b), respectively. Without error, we achieved approximation ratios of around $0.42$ with optimal and $0.39$ with original initial values. The approximation ratio is strongly affected by bit-phase flip errors, like the fidelity in Fig.~\ref{fig:noise_simulation_fid}, and has values of around $0.33$ and $0.30$ with optimal and original initial values at $\lambda = 1\%$. QAOA under noise results in a better approximation ratio and a significantly better success probability when optimal (rather than original) initial values are used. The influence of the type of initial values on approximation ratio is smaller than success probability. We conclude from the simulation results that as the error rate increases, the fidelity decreases, leading to a lower approximation rate and a slight decrease in success probability.

\section{Calibration-Aware Transpilation Procedure}
\label{sec:ca}

As has been discussed above (Fig.~\ref{fig:flow}), calibration-aware (CA) transpilation is organized in three steps: 
\begin{itemize}
    \item Topology-Aware Pre-Transpilation (TAPT), which can be computationally complex and produces a high-quality (or even optimal) solution that is independent of error rates.
    \item Noise-Aware Matching (NAM), which takes the connectivity determined during the first step and maps it to a sub-graph of the IBM QX's topology graph with the lowest error rates. This step is simple and needs to be repeated only if the error rates according to the calibration data have changed significantly.
    \item Decomposition and Optimization (DO), which is a collection of inexpensive procedures that take the ansatz circuit's parameters (for QAOA: angle $\theta_i$) into account. For example, certain quantum gates can be removed altogether for $\theta_i = 0$.
\end{itemize}

In the following, we provide details on the three steps.

\subsection{Topology-Aware Pre-Transpilation (TAPT)}

\begin{figure*}[htbp]
	\centerline{\includegraphics[width=\linewidth]{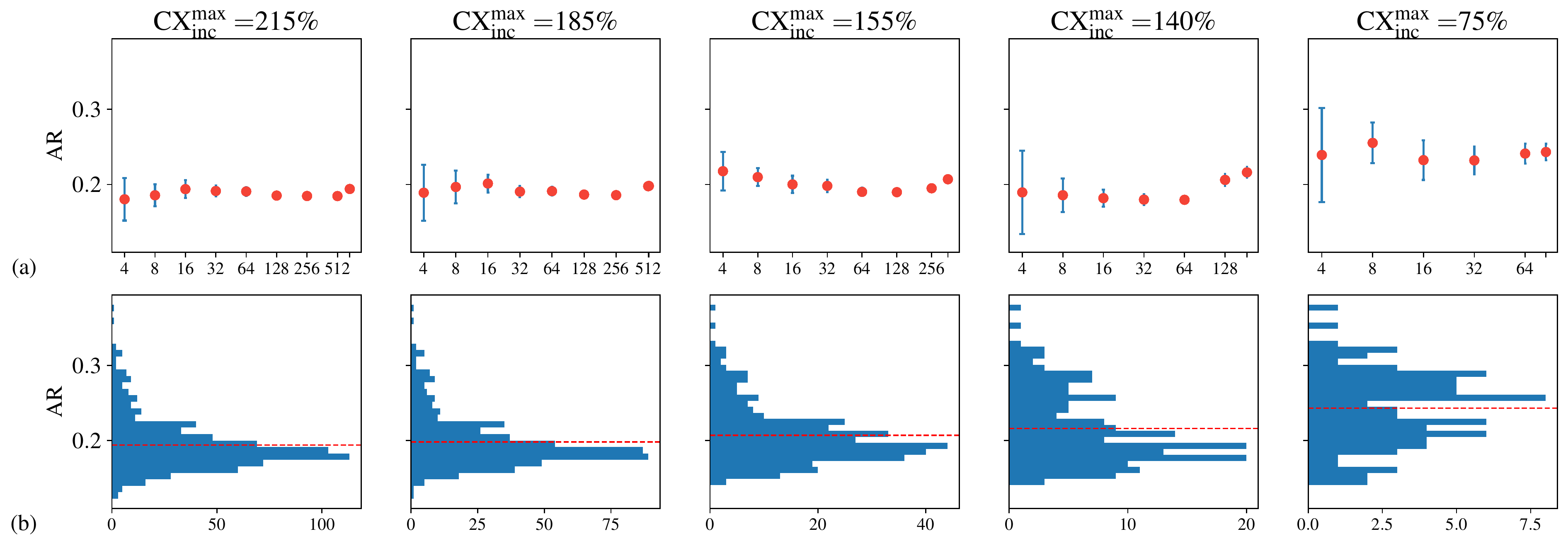}}
	\caption{$95\%$ confidence interval (a) and histogram (b) for approximation ratio of 678 QAOA repetitions using Qiskit transpiler on {\em ibmq\_ehningen}. With restriction of the amount of maximum increase in \texttt{cx} gates, the approximation ratio is improved.}
	\label{fig:conf_int_hist_ar}
\end{figure*}

\begin{figure*}[htbp]
	\centerline{\includegraphics[width=\linewidth]{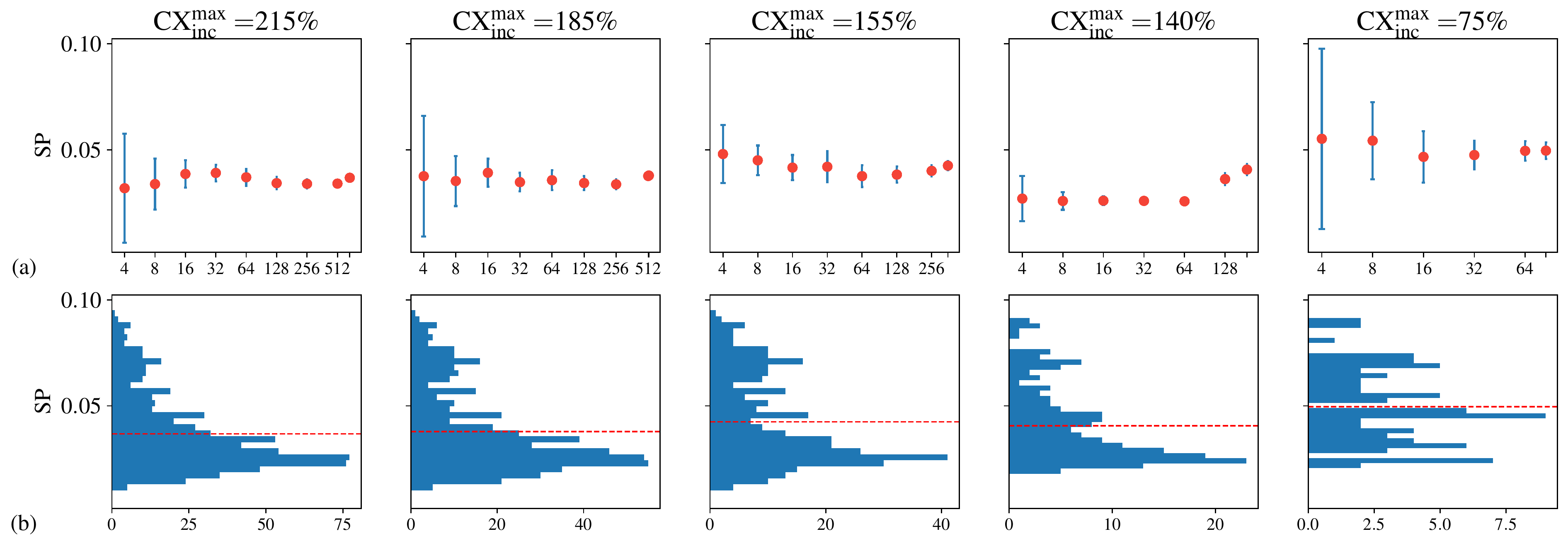}}
	\caption{$95\%$ confidence interval (a) and histogram (b) for success probability of $678$ QAOA repetitions using Qiskit transpiler on {\em ibmq\_ehningen}. Success probability shows a similar behavior to approximation ratio in Fig. \ref{fig:conf_int_hist_ar}.}
	\label{fig:conf_int_hist_pr}
\end{figure*}

Topology-aware pre-transpilation (TAPT) aims at satisfying the connectivity requirements of a quantum algorithm (here: ansatz circuit) at a given architecture. All two-qubit gates must either be mapped to connected qubits of the quantum hardware (e.g., qubits 10 and 12 in Fig.~\ref{fig:error_map}), or additional \texttt{swap} gates must be inserted such as to bring them onto neighboring qubits. On IBM's architecture used in this work, \texttt{swap} gates are rather expensive primitives, implemented by $3$ \texttt{cx} gates. Therefore, TAPT aims at minimizing the number of required extra \texttt{swap} gates, and ultimately the total number of \texttt{cx} gates in the circuit. Note that in general, even an optimized transpiled circuit has more \texttt{cx} gates than before transpilation.

Qiskit's transpilation procedure includes randomization, and running it multiple times produces different solutions. To improve stability, we implemented an additional check that bounds the maximum increase of \texttt{cx} gates $\texttt{CX}_{\rm inc}^{\max}$, i.e., discards transpiled circuits with an increase of \texttt{cx} gates exceeding this threshold.
Fig.~\ref{fig:conf_int_hist_ar} and Fig.~\ref{fig:conf_int_hist_pr} show the approximation ratio and, respectively, the success probability of 678 repetitions of QAOA on the physical quantum computer {\em ibmq\_ehningen} with $\texttt{CX}_{\rm inc}^{\max}$ set to $215\%, 185\%, 155\%, 140\%, 75\%$. For each such restriction, Fig.~\ref{fig:conf_int_hist_ar}a and Fig.~\ref{fig:conf_int_hist_pr}a show the $95\%$ confidence interval of approximation ratio and success probability, whereas Fig.~\ref{fig:conf_int_hist_ar}b and Fig.~\ref{fig:conf_int_hist_pr}b include the complete histograms. It can be seen that restricting the increase in \texttt{cx} gates tends to improve the performance of QAOA. Therefore, one main goal of our calibration-aware transpilation is to minimize the number of used \texttt{cx} gates (while also improving fidelity).

\RestyleAlgo{ruled}
\begin{algorithm}
	\caption{Topology-Aware Pre-Transpilation (TAPT)}\label{alg:calibration_transp}
	\KwInput{Original quantum circuit $qc$, Coupling map $G(V,E)$}
	\KwOutput{Topology-aware pre-transpiled circuit $pqc$}
	\Begin{
		$M$ $\gets$ initial mapping\;
		$U \gets$ set of sub-circuits between two qubits in {\em qc} with different structures\;
		\For{$u \in U$}{transform all gates in $qc$ with the same structure as $u$  to $u$ with gate parameters\;}
		Set initial mapping $M$\;
		Route the re-constructed circuit by inserting \texttt{swap} gates using SMT based optimal algorithm to minimize depth\;
		Decompose parameterized $u$ into basis gates of $qc$\; \For{\rm each \texttt{swap}
		gate in the circuit}{\eIf{\rm \texttt{swap} gate 
		is before the measurement}{
		remove \texttt{swap} gate and interchange the measurement of two qubits\;	}{
		decompose \texttt{swap} into three \texttt{cx} gates and optimize with \texttt{cx} cancellation\;
		}}
		Transform into a logic circuit by removing the idle wires\;
		\Return $pqc$
		}
\end{algorithm}

Algorithm \ref{alg:calibration_transp} shows the pseudocode of topology-aware pre-transpilation (TAPT). To obtain an efficient pre-transpiled circuit as a starting point, the algorithm starts with finding an initial mapping by {\em graph placement} \cite{Sivarajah2020tketAR}. This procedure identifies a sub-graph isomorphism between the graph of interacting logical qubits and the connectivity graph of the physical qubits. 
Before inserting \texttt{swap} gates, the circuit is partitioned into sub-circuits bounded by a \texttt{cx} gate on one or both sides. For QAOA ansatz circuits, such efficient sub-circuits have the form \texttt{cx rz cx} and implement the $Z\otimes Z$ interaction between two qubits.

To insert \texttt{swap} gates, we use an optimal method based on SMT (satisfiability modulo theory) \cite{TanC20} with circuit depth as the optimization objective to guarantee a high quality of pre-transpilation, as its runtime does not influence the performance of total process. The SMT method treats the sub-circuits identified as outlined above as primitive circuit elements. That is, \texttt{swap} gates are inserted only between sub-circuits. The rationale behind this procedure is the later application of \texttt{cx} cancellation, where \texttt{cx} gates on the boundaries of sub-circuits can be merged with \texttt{cx} gates that implement \texttt{swap} gates. In addition, considering sub-circuits reduces the problem complexity and the run-time of the SMT solver. Thereafter, the sub-circuits and the inserted \texttt{swap} gates are decomposed into basis gates of the circuit and undergo optimization, including \texttt{cx} cancellation. The resulting circuit is depth-optimized and executable on (a sub-graph of) the target topology graph.

\subsection{Noise-Aware Matching (NAM)}

The previous step maps the algorithm's qubits to a sub-graph of the topology graph, but it does not consider error rates of physical qubits in that sub-graph. At the same time, most topology graphs of today's larger-scale quantum computers have a large number of isomorphies and symmetries. For example, the topology graph from Fig.~\ref{fig:error_map} can be understood as consisting of two ``tiles'' (physical qubits 0, \ldots, 14, 16 and physical qubits 10, 12, \ldots, 26). Any algorithm mapped to a sub-graph from one ``tile'' can also run on the other. Moreover, each ``tile'' is symmetric. For instance, a 5-qubit algorithm mapped to physical qubits (0, 1, 4, 7, 6) is automatically valid for physical qubits (10, 12, 15, 18, 17); (15, 12, 10, 7, 6); (16, 14, 11, 8, 9); (26, 25, 22, 19, 20). Note that larger IBM computers have even more identical ``tiles'' and offer more valid alternatives for each result of step TAPT.

Noise-Aware Matching (NAM) considers up to $N$ alternative sub-graphs and selects the one with the highest effective average fidelity. $N$ is a user-defined constant, which trades the likelihood of finding a good matching against the number of necessary computations; the latter can be important when calibration-aware transpilation is used in the incremental mode and the quantum computer waits until the new sub-graph is identified. The effective average fidelity of quantum circuit $qc$ with the calibration data $\xi = (f_{\texttt{u}}, f_{\texttt{cx}}, f_{\texttt{d}})$ as:

\begin{equation}
\text{\em get\_fidelity}(qc, \xi) = \frac{1}{3}(\prod_{\texttt{u}\in qc} f_{\texttt{u}} + \prod_{\texttt{cx}\in qc} f_{\texttt{cx}} + \prod_{\texttt{d}\in qc} f_{\texttt{d}})
\end{equation}
where $f_{\texttt{u}}$, $f_{\texttt{cx}}$ and $f_{\texttt{d}}$ are fidelities of single qubit, \texttt{cx} gate and readout gate, respectively.

\begin{algorithm}
	\caption{Noise-Aware Matching (NAM)}\label{alg:noise_aware_matching}
	\KwInput{Topology-aware pre-transpiled circuit $pqc$, Coupling map $G(V,E)$, Calibration data $\xi$, Fidelity computation function {\em get\_fidelity}, Number of trials $N$}
	\KwOutput{Selected physical qubits}
	\Begin{
		$mqc \gets $ matched {\em pqc} with Qiskit\;
		$c_m \gets $ number of \texttt{cx} gates in $mqc$\;
		$f_m \gets $ {\em get\_fidelity}$\left(mqc, \xi\right)$\;
		$j \gets 0$\;
		\While{$j \neq N$}{
			$rqc \gets $ re-matched {\em pqc} with Qiskit\;
			$c_r \gets $ number of \texttt{cx} gates in $rqc$\;
			$f_r \gets $ {\em get\_fidelity}$\left(rqc, \xi\right)$\;
			\If{$c_r \leq c_m $ and $f_r > f_m$}{$mqc \gets rqc$}
			$j \gets j+1$\;
		}
	\Return Physical qubits used in $mqc$
	}
\end{algorithm}

\begin{figure*}[htbp]
	\centerline{\includegraphics[width=\linewidth]{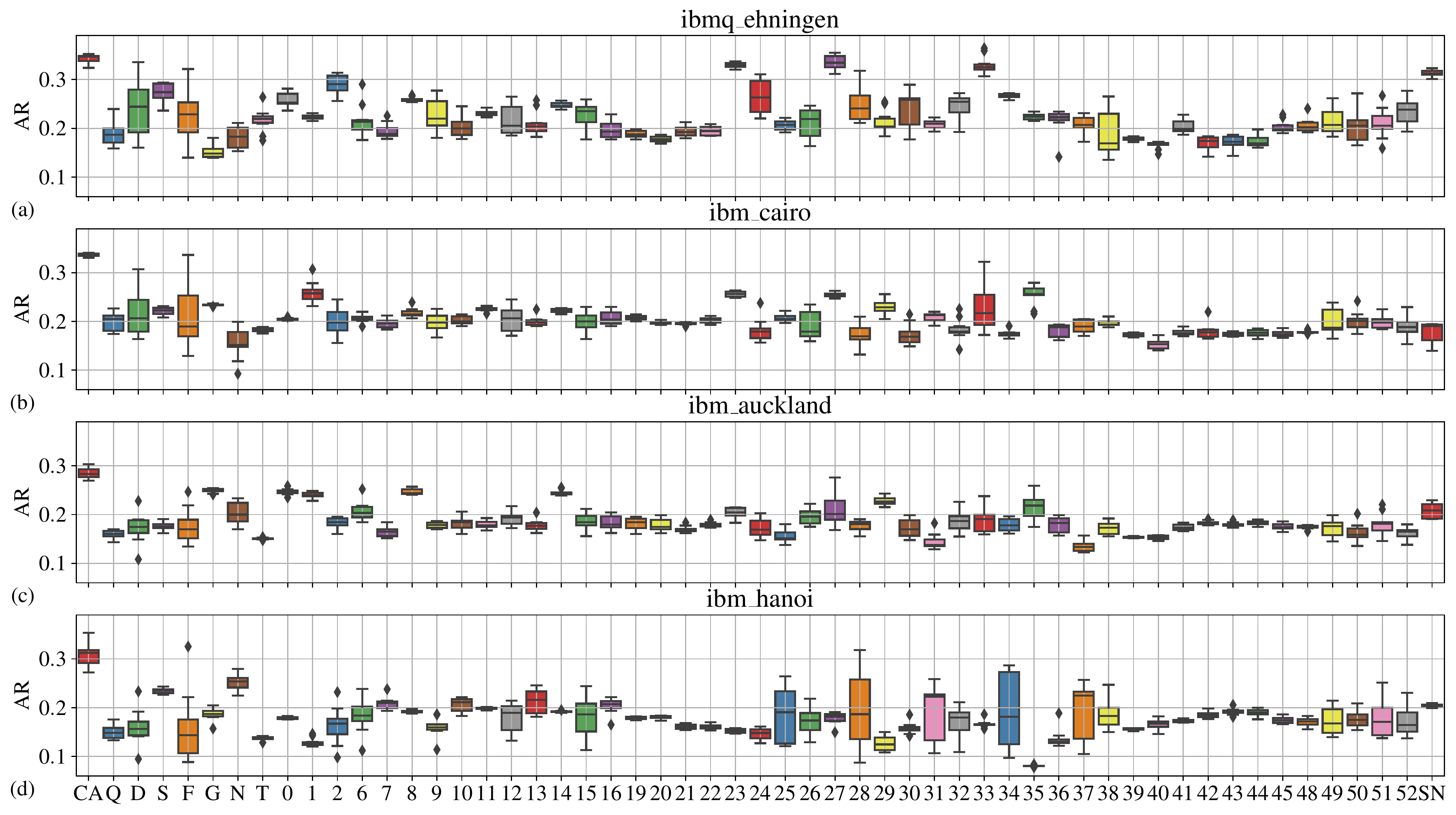}}
	\caption{Approximation ratio of 10 QAOA runs with different transpilation methods on four IBM QX computers. CA: Calibration-aware (this paper); Q: Qiskit's built-in transpilation procedure with {\em optimization\_level} 3; D/S/F: SMT-based methods \cite{TanC20} to minimize depth (D), minimize the number of \texttt{swap} gates (S), to maximize fidelity (F); G: t$\vert {\rm ket} \rangle$ with initial mapping based on {\em graph placement} \cite{Sivarajah2020tketAR}; N: t$\vert {\rm ket} \rangle$ with {\em noise aware placement} \cite{Sivarajah2020tketAR}; T: staq \cite{amy2020staq}; 00\ldots52: methods composed of different initial mapping and routing procedures, some including ZX-calculus optimization \cite{kissinger2020Pyzx}; SN: swap network \cite{harrigan2021quantum}. 10 runs of QAOA per data point.}
	\label{fig:ar_all_methods}
\end{figure*}

The noise-aware matching (NAM) is described in Algorithm \ref{alg:noise_aware_matching}. As input we have pre-transpiled circuit, which has satisfied the connectivity of sub-graph of topology graph, coupling map, the latest calibration data, the function {\em get\_fidelity} to calculate the fidelity of circuit, and the number of trials $N$ (we used $N = 15$ in our experiments). In order to select high-fidelity qubits, we perform $N$ trial matchings with Qiskit's transpile. With the randomization of Qiskit's transpilation procedure, the pre-transpiled circuit is matched to different physical qubits of quantum computer. Effective average fidelities of the matched circuits on physical qubits are calculated and the circuit with the highest fidelity is picked. This circuit contains the information of selected physical qubits in $N$ trials.

\subsection{Decomposition and Optimization (DO)}

After the NAM process, the target qubits used to run the algorithm are fixed. Then the quantum algorithm is decomposed into the native gate set supported by IBM QX. In this work, we are using IBM's computers that support single-qubit gates and the \texttt{cx} gate as an entangling gate. After decomposition, we apply optimization techniques {\em Optimize1qGates}, {\em CommutationAnalysis}, {\em CommutativeCancellation}, {\em CXCancellation}, {\em RemoveDiagonalGatesBeforeMeasure} and {\em RemoveResetInZeroState} provided by Qiskit. We repeat this process $15$ times aiming to obtain the final transpiled circuit with the least number of \texttt{cx} gates. Note that this step is architecture-specific and would need to be adapted for a different platform; for example, Google's computers use \texttt{cz} gates as entangled gates.

\section{Benchmarking with QAOA}
\label{sec:expres}

In this section, we benchmark the calibration-aware (CA) transpilation with QAOA on four IBM QX computers, {\em ibmq\_ehningen}, {\em ibm\_cairo}, {\em ibm\_auckland} and {\em ibm\_hanoi}, all of which have $27$ qubits and the same topology graph, as shown in Fig.~\ref{fig:error_map}. The initial values of QAOA used here are determined by COBYLA, i.e. the original values labeled in Fig.~\ref{fig:noise_simulation}. We use three sets of metrics: the quality of the transpilation process (quantified by percental increase of the circuit's depth $\Delta d\%$, its total number of gates $\Delta g\%$ and number of its \texttt{cx} gates $\Delta g_{\texttt{cx}}\%$ as a result of transpilation);
the runtime of the transpilation procedure; and the quality of QAOA in terms of approximation ratio and success probability achieved on a physical quantum computer.

We start with a comparison of CA with a large number of different transpilation methods with respect to approximation ratio and success probability. Then, we pick one of the best methods observed and compare it with CA in-depth. Finally, we outline the potential benefits of CA for runtime of a complete QAOA algorithm.

\subsection{Approximation Ratio, Success Probability, \texttt{cx} Gate Count}

\begin{figure*}[htbp]
	\centerline{\includegraphics[width=\linewidth]{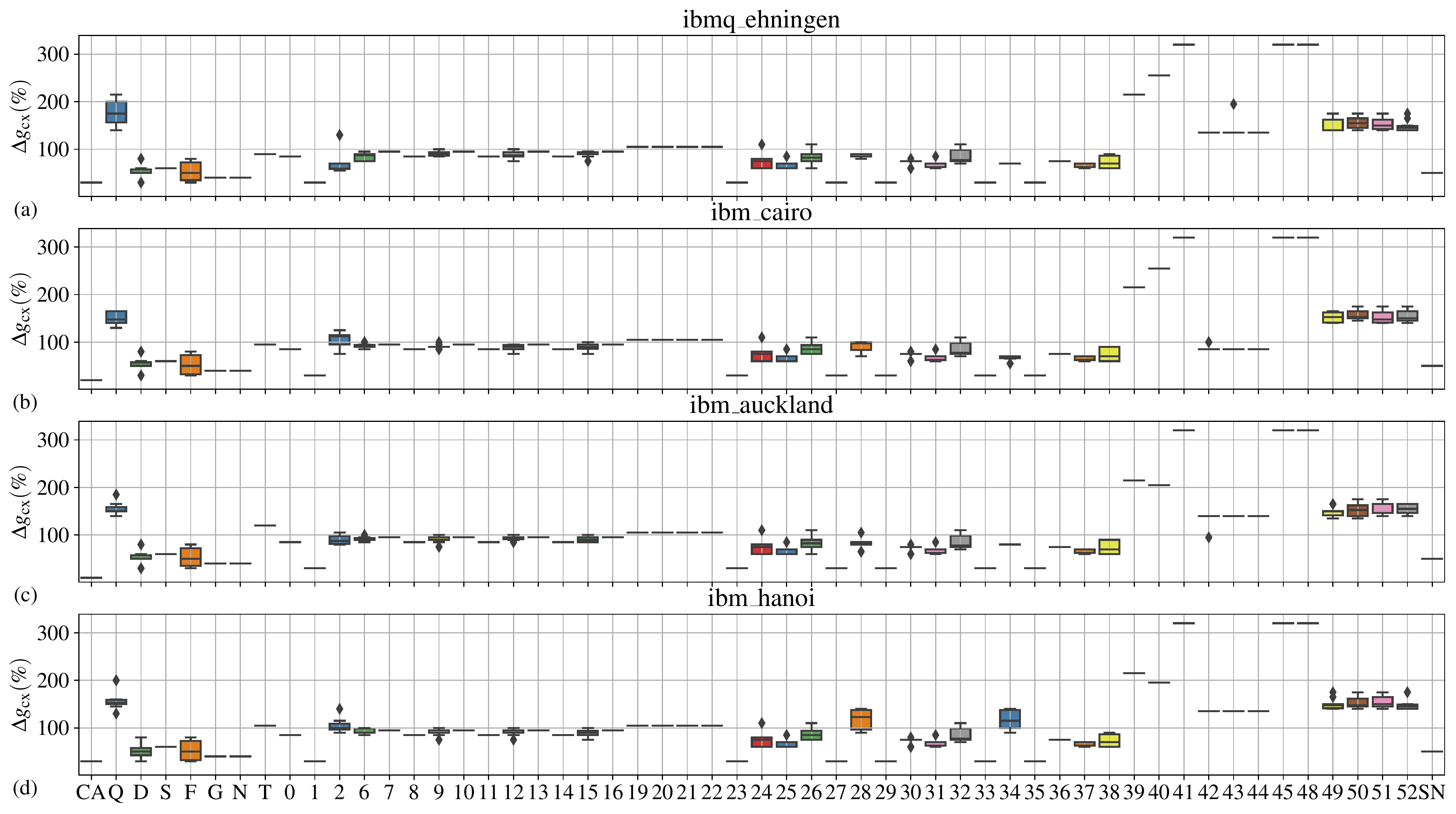}}
	\caption{Percental increase in the number of \texttt{cx} gates after transpilation (10 QAOA runs, same methods as in Fig.~\ref{fig:ar_all_methods}).}
	\label{fig:cx_all_methods}
\end{figure*}

We implemented a number of transpilation methods and compared the achieved approximation ratio in Fig.~\ref{fig:ar_all_methods}. In addition to CA, this figure includes Qiskit's built-in transpilation procedure, SMT-based methods \cite{TanC20}, two variants of t$\vert {\rm ket} \rangle$ \cite{Sivarajah2020tketAR}, staq \cite{amy2020staq}, a total of 53 composite methods depending on different initial mapping and routing procedures, and swap network (SN) \cite{harrigan2021quantum}. CA either outperforms other methods or is on a par with the best of them for all four quantum computers.

\begin{table*}[t!]
	\caption{Comparison of percentage increase in depth $\Delta d\%$, the total number of gates $\Delta g\%$ and the number of \texttt{cx} gates $\Delta g_{\texttt{cx}}\%$ after transpilation with CA and SF. $\mu$: average value. $\sigma$: standard deviation.}
	\begin{center}
		\begin{tabular}{|c|c|c|c|c|c|c|c|c|c|c|c|c|}
			
			\hline
			\textbf{}&\multicolumn{4}{|c|}{\textbf{$\Delta d \%$} }& \multicolumn{4}{|c|}{\textbf{$\Delta g \%$} } & \multicolumn{4}{|c|}{\textbf{$\Delta g_{\texttt{cx}}\%$} } \\
			\cline{2-13}
			
			\textbf{}&\multicolumn{2}{|c|}{\textbf{CA} }& \multicolumn{2}{|c|}{\textbf{SF} }&\multicolumn{2}{|c|}{\textbf{CA} }& \multicolumn{2}{|c|}{\textbf{SF} }&\multicolumn{2}{|c|}{\textbf{CA} }& \multicolumn{2}{|c|}{\textbf{SF} } \\
			\cline{2-13} 
			\textbf{IBM QX} &  \textbf{$\mu$}& \textbf{$\sigma$} & \textbf{$\mu$}& \textbf{$\sigma$}&  \textbf{$\mu$}& \textbf{$\sigma$} & \textbf{$\mu$}& \textbf{$\sigma$}&  \textbf{$\mu$}& \textbf{$\sigma$} & \textbf{$\mu$}& \textbf{$\sigma$} \\
			\hline
			\textit{ibmq\_ehningen}& 126.32 &         0.0 &  116.84 &         31.47 &  170.00 &         0.0 &  104.00 &         66.78 &   30.00 &         0.0 &   42.00 &         17.20 \\
			\hline
			\textit{ibm\_cairo}     &       126.32 &         0.0 &  110.00 &         27.49&  170.00 &         0.0 &   93.20 &         63.03&   30.00 &         0.0 &   48.00 &         18.87 \\
			\hline
			\textit{ibm\_auckland}  &       126.32 &         0.0 &  116.84 &         31.47&  170.00 &         0.0 &  104.00 &         66.78&   30.00 &         0.0 &   42.00 &         17.20\\
			\hline
			\textit{ibm\_hanoi}     &       126.32 &         0.0 &  116.84 &         31.47 &  170.00 &         0.0 &  104.00 &         66.78&   30.00 &         0.0 &   42.00 &         17.20 \\
			\hline
		\end{tabular}
		\label{tab:circ_properties}
	\end{center}
\end{table*}

We believe that this advantage is due to CA's NAM step considering a number of possible sub-graphs, selecting the one with the best fidelity according to a more up-to-data calibration data than other methods. At the same time, the TAPT step produces a robust depth-optimized solution for the connectivity constraints, thus limiting errors that stem from excessive gate-count. We observed that purely fidelity-oriented transpilation can incur  strong variability in gate count for different ansatz circuits $A(\theta_1), A(\theta_2), \ldots$; CA's DO step is applying only minimal modifications to the basic solution from the TAPT step, thus leading to well-aligned transpilation results for different circuits.


The percental increase in \texttt{cx} gates after transpilation is reported in Fig.~\ref{fig:cx_all_methods}. The numbers differ only minimally among the four quantum computers. Again, CA is consistently best or among the best methods with respect to this metric, while other methods produce an increase of up to $330\%$ (more than four times) in \texttt{cx} gates. CA's outcome is also much more stable, since all its ansatz circuits are based on the same high-quality basic solution provided by the TAPT step, whereas, e.g., Qiskit (Q) produces quite different transpilation results for each QAOA run and quantum computer.

Table \ref{tab:circ_properties} shows a detailed comparison of CA with SF in terms of increase in depth $\Delta d\%$, total number of gates $\Delta g\%$ and number of \texttt{cx} gates $\Delta g_{\texttt{cx}}\%$. The table shows that SF is a good transpilation method with a slightly better $\Delta d\%$, significantly better $\Delta g\%$, but significantly worse $\Delta g_{\texttt{cx}}\%$. SF exposes large-scale variability whereas CA's results are repeatable.

\begin{figure}[htbp]
	\centerline{\includegraphics[width=\linewidth]{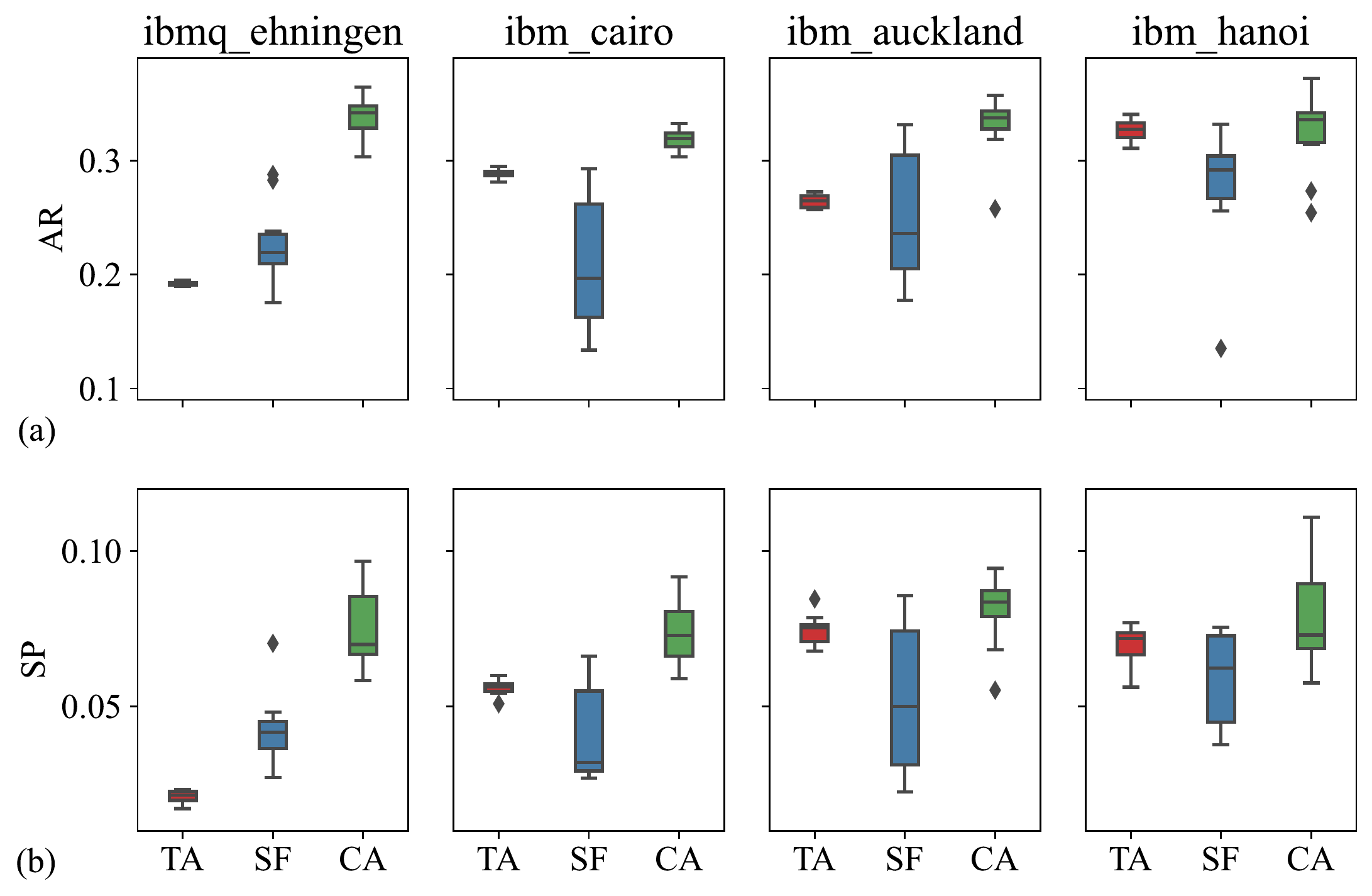}}
	\caption{Approximation ratio and success probability with IBM QX computers. TA: topology-aware (noise-unaware) transpilation (TAPT step of CA); SF: SMT-based fidelity maximization; CA: full calibration-aware method (TAPT, NAM, DO).}
	\label{fig:ar_sp_tasfca}
\end{figure}

To assess the significance of calibration data and the NAM step, we report in Fig.~\ref{fig:ar_sp_tasfca} the approximation ratio and the success probability of three methods: topology-aware transpilation TA, which is CA that stopped after the TAPT step and did not incorporate any calibration data; SMT-based transpilation SF \cite{TanC20} that maximizes the circuit's fidelity and is used for reference; and the full CA procedure with all its three steps. CA by far outperforms TA and is also consistently better than SF, while both TA and CA are less affected by the variability of the obtained results. We conclude that all three steps of CA are needed to obtain a high-quality solution.

\subsection{Runtime}

One central objective of our calibration-aware (CA) transpilation approach is to reduce the runtime of iterative variational algorithms.
In this section, we compare CA with SMT-based method in \cite{TanC20} that maximizes the circuit's fidelity (SF).
Table \ref{tab:runtime} reports the average runtimes of CA's three steps, their sum ($\mu$) and standard deviation ($\sigma$), along with the average runtime and standard deviation for SF, which is not partitioned into steps. It can be seen that the overall runtimes of CA and SF are comparable, suggesting a similar amount of computational effort being invested. However, CA manages to serialize its computation without a major deterioration of overall runtime. Furthermore, the standard deviation of total runtime with CA is an order of magnitude smaller than with SF, which means less latency due to transpilation.
\begin{table}[tb]
	\caption{Comparison of runtime of transpilation with CA and SF for 10 QAOA runs (in seconds). $\mu$: average runtime. $\sigma$: standard deviation.}
	\begin{center}
		\begin{tabular}{|c|c@{\;\;}|@{\;\;}c@{\;\;}|@{\;\;}c@{\;\;}|@{\;\;}c@{\;\;}|@{\;\;}c|c|c|}
			\hline
			\textbf{}&\multicolumn{5}{|c|}{\textbf{CA} }& \multicolumn{2}{|c|}{\textbf{SF} } \\
			\cline{2-8} 
			\textbf{IBM QX} & \textbf{TAPT}& \textbf{NAM}& \textbf{DO} & \textbf{$\mu$}& \textbf{$\sigma$} & \textbf{$\mu$}& \textbf{$\sigma$} \\
			\hline
			\textit{ibmq\_ehningen}& 18.13 &      4.41 &     2.26 &  24.80 &     0.22 &  30.00 &          3.65 \\
			\hline
			\textit{ibm\_cairo}     &       23.07 &      3.59 &     3.13 &  29.79 &     0.26 &  25.90 &          1.74 \\
			\hline
			\textit{ibm\_auckland}  &       22.92 &      3.44 &     2.31 &  28.67 &     0.14 &  27.94 &          5.46 \\
			\hline
			\textit{ibm\_hanoi}     &       17.44 &      4.37 &     2.50 &  24.31 &     0.31 &  29.89 &          3.20 \\
			\hline
		\end{tabular}
		\label{tab:runtime}
	\end{center}
\end{table}

To illustrate the runtime advantage enabled by CA, 
assume we have $N_A$ QAOA ansatz circuits to execute. The expectation value of the total runtime for $N_A$ circuits with SF is
\begin{equation}
    \mu_{\rm SF}(N_A) = \mu_{\rm SF}\times N_A
\end{equation}
with $\mu_{\rm SF}$ being the average runtime for one circuit. If the calibration data changes every $m$ iterations, the expected total runtime with CA is
\begin{equation}
    \mu_{\rm CA}(N_A) = \mu_{\rm TAPT} + \mu_{\rm NAM}\times\ceil*{\frac{N_A}{m}}+\mu_{\rm DO}\times N_A
\end{equation}
where $\mu_{\rm TAPT}$, $\mu_{\rm NAM}$ and $\mu_{\rm DO}$ are average runtimes for TAPT, NAM and DO process, respectively.
For the special case that calibration data is fixed, we have $m = N_A$ and only the last step DO needs to be executed each time.

\begin{table}[tb]
	\caption{Comparison of runtime (in seconds) of CA and SF for $N_A = 5, 100$. CER: Changing error rate after $m=5$ iterations. FER: Fixed error rate ($m=N_A$). $\Delta \mu$: average time savings compared to SF.}
	\begin{center}
		\begin{tabular}{|@{\;\;}c@{\;\;}|@{\;\;}c@{\;\;}|c|c|c|c|c|}
			\hline	\textbf{}&\textbf{}&\multicolumn{2}{|c|}{\textbf{CA} }& \multicolumn{1}{|c|}{\textbf{SF} }& \multicolumn{2}{|c|}{\textbf{$\Delta \mu (\%)$} } \\
			\cline{3-7}
			$N_A$&\textbf{IBM QX} & \textbf{CER}& \textbf{FER}& \textbf{}& \textbf{CER}& \textbf{FER} \\
			\hline
			&\textit{ibmq\_ehningen} &       33.83 &         33.83 &  149.98 &        -77.44 &          -77.44 \\
			\cline{2-7}
			$5$&\textit{ibm\_auckland}  &       37.91 &         37.91 &  139.71 &        -72.87 &          -72.87 \\
			\cline{2-7}
			&\textit{ibm\_cairo}     &       42.33 &         42.33 &  129.48 &        -67.31 &          -67.31 \\
			\cline{2-7}
			&\textit{ibm\_hanoi}     &       34.32 &         34.32 &  149.43 &        -77.03 &          -77.03 \\
			\hline
			
			&\textit{ibmq\_ehningen} &       332.23 &          248.46 &  2999.66 &        -88.92 &          -91.72 \\
			\cline{2-7}
			$100$&\textit{ibm\_auckland}  &        322.67 &          257.31 &  2794.29 &        -88.45 &          -90.79 \\
			\cline{2-7}
			&\textit{ibm\_cairo}     &       408.21 &          339.99 &  2589.61 &        -84.24 &          -86.87\\
			\cline{2-7}
			&\textit{ibm\_hanoi}     &       355.05 &          272.03 &  2988.57 &        -88.12 &          -90.90 \\
			\hline
		\end{tabular}
		\label{tab:runtime_hyp}
	\end{center}
\end{table}

The projected runtimes of a complete QAOA run with $N_A \in \{5, 100\}$ ansatz circuits are shown in Table \ref{tab:runtime_hyp}. The data assumes two scenarios: changing error rate (CER), where the calibration data changes after $5$ iterations ($m = 5)$, and fixed error rate (FER), where the calibration data remains unchanged ($m = N_A$). We see an improvement of up to one order of magnitude due to CA's three-step structure where the most expensive part of the calculation is executed only once.

\subsection{Conclusion and Comparison}

For evaluation, we compared CA approach with several other methods. The experiments show that applying our approach yields better and more stable results. The main advantage of CA is that TAPT needs to be performed only once and a number of trials in the fast NAM processing qualify that the implementation is performed on qubits with high fidelity. Another highlight is that the run time for $N_A$ ansatz circuits is significantly reduced: up to $88.92\%$ with CER and $91.72\%$ with FER for $N_A = 100$. Moreover, with CA we have stable properties for transpiled circuit, which are shown by the increase of depth, number of gates and number of \texttt{cx} gates. All this guarantees that CA produces consistently high quality on four IBM quantum computers.

\section{Conclusions and Future Work}
\label{sec:concl}

The decisive role of variational algorithms during the NISQ era justifies a specialized transpilation approach for such algorithms. Calibration-aware transpilation leverages the knowledge that subsequent ansatz circuits have the same basic structure and differ only in their parameters. It naturally adapts itself to abrupt changes in the error rates of the quantum computer executing the algorithm, which is a reality today. Our results show that calibration-aware transpilation strikes a good balance between quality and stability of transpilation results and saves time thanks to offloading the heaviest computation to a procedure that is run one time for all ansatz circuits.

Our findings are confirmed by results for QAOA obtained on four physical quantum computers with a similar architecture. They are compared with an extensive set of previous transpilation procedures executed on the same computers and are put into perspective with simulations assuming standard error models. We believe that calibration-aware transpilation is particularly attractive for today's quantum cloud computers with their potentially long queuing times: if a circuit is executed long after it has been transpiled, its calibration data can become outdated and the actual error rate can get much worse than expected. Calibration-aware transpiliation enables incremental operation, where the execution starts almost instantly after calibration, with only lightweight parts of the transpilation procedure being 
completed in between.

For the future, we are interested in further improving the algorithm's performance, especially for emerging NISQ computers with 100s or 1000s qubits. Here, quick variants of the NAM step that leverage the architecture's symmetries are essential. Moreover, the TAPT step must be further evaluated for stability for other variational algorithms and quantum architectures. Another interesting question is whether we can make calibration-aware transpilation provably optimal with respect to one of the targets, despite being divided into three independent steps. This would make optimal approaches available for time-critical quantum circuit execution, as the expensive TAPT step can be performed on a classical computer before any access to a quantum computer.

\bibliographystyle{IEEEtran}
\bibliography{qce-biblio}

\end{document}